\documentclass{article}

\usepackage{PRIMEarxiv}

\usepackage[utf8]{inputenc} 
\usepackage[T1]{fontenc}    
\usepackage{hyperref}       
\usepackage{url}            
\usepackage{booktabs}       
\usepackage{amsfonts}       
\usepackage{microtype}      
\usepackage{lipsum}
\usepackage{fancyhdr}       
\usepackage{graphicx}       
\graphicspath{{media/}}     

\title{Parameters identification for an inverse problem arising from a binary option using a Bayesian inference approach}

\author{
  Yasushi Ota \\
  Faculty of Business Administration\\
  St. Andrew's University\\
  1-1, Manabino, Izumi City, Osaka, Japan\\
  \texttt{yasushio@andrew.ac.jp} \\
   \And
  Yu Jiang \\
  School of Mathematics\\
  Shanghai University of Finance and Economics \\
  777 Guoding Rd. Shanghai,200433, P. R. China\\
  \texttt{jiang.yu@mail.shufe.edu.cn} \\
  \And
  Daiki Maki \\
  Faculty of Commerce\\
  Doshisha University\\ Karasuma-higashi-iru, Imadegawa-dori, Kamigyo-ku, Kyoto, Japan\\
  \texttt{dmaki@mail.doshisha.ac.jp} \\
}

\begin{document}
\maketitle

\begin{abstract}
No--arbitrage property provides a simple method for pricing financial derivatives. However, arbitrage opportunities exist among different markets in various fields, even for a very short time. 
By knowing that an arbitrage property exists, we can adopt a financial trading strategy.
This paper investigates the inverse option problems (IOP) in the extended Black--Scholes model. 
We identify the model coefficients from the measured data and attempt to find arbitrage opportunities in different financial markets using a Bayesian inference approach, which is presented as an IOP solution.
The posterior probability density function of the parameters is computed 
from the measured data.
The statistics of the unknown parameters are estimated by a Markov Chain Monte Carlo (MCMC) algorithm, which exploits the posterior state space. 
The efficient sampling strategy of the MCMC algorithm enables us to solve inverse problems by the Bayesian inference technique. 
Our numerical results indicate that the Bayesian inference approach can simultaneously estimate the unknown trend and volatility coefficients from the measured data.

\end{abstract}


\section{Introduction}
he technique of inverse problems 
for a partial differential equation of a parabolic type 
is developed and used in various fields, 
such as Inverse heat transfer problems(IHTP), Inverse heat conduction problems(IHCP), Inverse option problems(IOP), 
etc\cite{Alif, beck, B-I, dupire}.

In this paper we consider the backward parabolic equation:
\begin{equation}
\left\{
\begin{array}{cl}
\displaystyle\frac{\partial u}{\partial t} + \frac{1}{2} \sigma(x,t)^2 x^2 \frac{\partial^2 u}{\partial x^2} + \mu(x,t) x \frac{\partial u}{\partial x} - ru =0, \hspace{1ex} (x,t) \in (0,\infty)\times[0, T), \\
\mbox{} \\
\displaystyle u(x,t){|}_{t=T} = \Phi(x,T), \hspace{8ex} x \in (0,\infty).
\end{array}
\right.
\label{maineq1}
\end{equation}
where $u(x,t)$ is the price for a derivative, 
such as an option, bond, interest rate, futures, foreign exchange, etc.
Moreover, $x$ in the underlying asset price, $t$ is the time, 
$\mu(x,t)$ and $\sigma(x,t)$ are the drift and volatility coefficient of the process $x$, the interest rate $r$ is a nonnegative constant, 
and $K$ is the strike price and $T$ is the maturity of the underlying asset, and $\Phi(x,T)$ is a suitable initial condition. 

Now, we are interested in the following inverse option problem(IOP):
Let the current time $t^*$ be given , and determine simultaneously $\mu(x,t)$ and $\sigma(x,t)$ from the observation of data $u(x,t^*), x \in \omega,$
where $\omega$ is the interval.

IOP in mathematical finance 
were started by Dupire \cite{dupire}.
He derived the option premium 
$U(T, K)$ as a solution $u(\cdot, \cdot ; T, K)$ 
to the dual equation of Black-Schoels equation, which is $\mu = r$ in (\ref{maineq1}), with respect to the strike price $K$ and maturity T as follows:
\begin{equation}
\displaystyle\frac{\partial U}{\partial T} - \frac{1}{2} \sigma(T, K)^2 K^2 \frac{\partial^2 U}{\partial K^2} + r K \frac{\partial U}{\partial K}=0. 
\label{dupire_eq}
\end{equation}
If the option price and its derivative can be determined for all possible 
$T$ and $K$, then the local volatility function $\sigma(T, K)$ can be directly derived from Eq.(\ref{dupire_eq}) as
\begin{equation}
\displaystyle \sigma(T, K)^2 = 
\displaystyle \frac{\displaystyle\frac{\partial U}{\partial T} + r K \frac{\partial U}{\partial K}}{\displaystyle \frac{1}{2} K^2 \frac{\partial^2 U}{\partial K^2}}.
\label{dupire_sol}
\end{equation}
Using this approach, we can deduce the local volatility function 
from the quoted option prices in the financial market. 
Bouchouev and Isakov \cite{B-I}, Bouchouev et al. \cite{B-I-V}, and Ota and Kaji \cite{yasukaji}, by using a linearization method, considered the following form of the time--independent local volatility function 
$\sigma(K)$:
\begin{equation}
\displaystyle\frac{1}{2} \sigma^2(K) = \frac{1}{2} \sigma_0^2 + f(K)
\end{equation}
where $f$ is a small perturbation of the constant volatility $\sigma_0$. 
Moreover, Mitsuhiro and Ota \cite{miyasu}, Korolev et al. \cite{kubo} and Doi and Ota \cite{Doi-yasu} used the extended Black--Scholes equation (\ref{maineq1}) and then reconstructed the drift function by linearization method.
The above studies provided point estimates of unknown parameters by exact determination or least squares optimization, without rigorously examining 
and considering the measurement errors in the inverse solutions.
In \cite{yasu_jiang_gen_ue}, we considered the Option Problem, 
which has an initial condition $\Phi(x, T)$ in (\ref{maineq1}) such as 
\begin{equation}
\Phi(x, T) =
\left\{\begin{array}{cc}
x-K & x \ge K \\
0 & x < K,
\end{array}\right.
\end{equation}
and simultaneously estimated the unknown drift and volatility coefficients from the measured data by applying Bayesian inference approach.

In this paper,  
we investigate the Binary Option Problem, which has an initial condition $\Phi(x,T)=H(x-K)$ in (\ref{maineq1}), 
where $H$ is the Heviside function, that is, 
\begin{equation}
H(x-K) =
\left\{
\begin{array}{cc}
1 & x \ge K \\
0 & x < K.
\end{array}
\right.
\end{equation}
And we attempt to derive the option pricing equation of so-called Dupire type which we were unable to derive in \cite{yasu_jiang_gen_ue}, and to simultaneously estimate the drift and volatility coefficients from the measured data. We can apply results of this study to the problem that we estimate the market risk from the pricing of derivatives such as interest rate.

Bayesian inference approach solves an inverse problem by formulating 
a complete probabilistic description of the unknowns and uncertainties from the given measured data (see \cite{mcmc-springer}).
Incorporating the likelihood function with a prior distribution, 
the Bayesian inference method provides the posterior probability density function (PPDF).
Owing to the recent developments in Bayesian inference work, 
including Bayesian inference approach by efficient sampling methods such as Markov Chain Monte Carlo (MCMC), 
we can apply the Bayesian inference technique to inverse problems in remote sensing \cite{Remote}, seismic inversion \cite{Seismic}, heat conduction problems \cite{Heat1}, \cite{Heat2} and various other real--world problems.
Moreover, several prior publications such as \cite{B-G-R,J-J,J-P1,T-Z,Tunaru} are related to option pricing based on Bayesian inference. 
In those publications, the option prices are usually computed by using the analytical solution (or so-called Black-Scholes formula) or applying of Monte Carlo simulation of original stochastic differential equation under an assumption which the volatility is constant. 

This paper is divided into five parts.
Our inverse problem is mathematically formulated in Section 2. 
Section 3 outlines the general Bayesian framework for solving inverse problems and discusses the numerical exploration of the posterior state space by the MCMC method. 
In Section 4, we discretize our inverse problem and reconstruct the parameters by a numerical algorithm.
We then discuss various aspects of our results through numerical examples. 
Concluding remarks are given in Section 5.

\section{Mathematical formulation of IOP}
\label{sec:mathematical formulation}

In this paper, 
we consider that the volatility is a constant ($\sigma(x,t)\equiv\sigma_0)$ and the initial condition is a step function in (\ref{maineq1}):
\begin{equation}
\left\{
\begin{array}{cl}
\displaystyle\frac{\partial u}{\partial t} + \frac{1}{2} \sigma_0^2 x^2 \frac{\partial^2 u}{\partial x^2} + \mu(x,t) x \frac{\partial u}{\partial x} - ru =0 \hspace{2ex} (x,t) \in (0,\infty)\times[0,T), \\
\mbox{} \\
\displaystyle u(x,t){|}_{t=T} = H(x-K) \hspace{5ex} x \in (0,\infty).
\end{array}
\right.
\label{E-BS-const}
\end{equation}

First, we check an idea of Dupire\cite{dupire}
and derive the partial differential equation dual to (\ref{E-BS-const}).

We set 
\begin{equation}
G(x,t; K,T) = -\frac{\partial u(x,t; K,T)}{\partial K}
\label{defofG}
\end{equation}
and then $G(S,t;K,T)$ satisfies the differential equation (\ref{E-BS-const}), and 
\begin{equation}
G(x,T; K,T) = \delta(x-K).
\end{equation}
According to Friedman\cite{Fri}, 
$G(x,t; K,T)$ satisfies for fixed $(x,t)$ as a function of $(K, T)$ the following differential equation and initial condition:
\begin{equation}
\left\{
\begin{array}{cl}
\displaystyle\frac{\partial G}{\partial T} - \frac{1}{2}\frac{\partial^2}{\partial K^2} 
(\sigma_0^2 K^2 G) + \frac{\partial}{\partial K}(\mu(K, T) K G) + rG =0 \hspace{1ex} (x,t) \in (0,\infty)\times[0,T), \\
\mbox{} \\
\displaystyle G(x,t;K,t){|}_{t=T} = \delta(x-K) \hspace{5ex} x \in (0,\infty).
\end{array}
\right.
\label{E-BS-dupire}
\end{equation}
Then, we use the definition of $G(S,T; K,T)$, and integrate the equation (\ref{E-BS-dupire}) from $K$ to $\infty$.
The third term in the left-hand side can be integrated by parts as follows
\begin{eqnarray*}
\int_K^{\infty} \frac{\partial}{\partial \xi}(\mu(\xi, T)\xi G) d\xi
=\mu(K,T) K \frac{\partial u}{\partial K}
\end{eqnarray*}
where we have used the following behaviour at infinity
\begin{equation}
u, K\frac{\partial u}{\partial K}, K^2 \frac{\partial^2 u}{\partial K^2} \to 0 \hspace{5ex} \mbox{as} \hspace{1ex} K \to \infty
\end{equation}

Consequently, we can obtain the following dual equation for $u(\cdot; K, T)$
\begin{eqnarray}
\displaystyle\frac{\partial u}{\partial T} - \frac{1}{2}\sigma_0 K^2 \frac{\partial^2 u}{\partial K^2} - (\sigma_0 - \mu(K, T)) K \frac{\partial u}{\partial K} + ru =0.
\end{eqnarray}

Now, the substitution 

$$\begin{array}{cc}
y = \log \displaystyle \frac{K}{x}, \hspace{5ex} \tau = T - t,\hspace{10ex} \\
\mu(y) = \mu(K, T), \hspace{5ex} U(y, \tau) = u(x, t; K, T) 
\end{array}
$$

transforms the equation and the initial condition  (\ref{E-BS-const}) into 
\begin{equation}
\left\{
\begin{array}{cl}
\displaystyle\frac{\partial U}{\partial \tau} - \frac{\sigma_0^2}{2} \frac{\partial^2 U}{\partial y^2} - \Bigl( \frac{\sigma_0^2}{2} - \mu(y) \Bigr)\frac{\partial U}{\partial y} + r U=0 \hspace{3ex} (y, \tau) \in \textbf{R} \times(0, \tau^*), \\
\mbox{} \\
U(y, 0) = H(-y) \hspace{10ex} y \in \textbf{R},
\mbox{} 
\end{array}
\right.
\label{E-BS-consider}
\end{equation}
where $\tau^* = T -t^*$ and $t^*$ is the current time.

Then, we consider the following problem IOP:

\noindent \textbf{Problem}

If we give the data $U_*(x) := U(y, \tau^*)$ on $\omega$ at $\tau=\tau_* = T-t^*$then identify $\sigma_0$ and $\mu(y)$ satisfying (\ref{E-BS-consider})

\vspace{3ex}

However, due to the nonlinearity of this inverse problem, 
the uniqueness and existence of its solution are hard to prove.
In this paper we attempts to reconstruct the parameters
by a statistical method simultaneously estimates $\mu(y)$ and $\sigma_0$ from the measured data $U^*(y)$. 

Let us define $m-$dimensional vectors $Y_{\bar{\theta}}$, $F(\theta)$ and $\varepsilon$ as follows:
\begin{eqnarray*}
\{Y_{\bar{\theta}}\}_{j} &=& U^*(y_j) = U(\tau^*, y_j;\bar{\theta})(1+\varepsilon_{j}) \\
\{F(\theta)\}_{j} &=& U(\tau^*, y_j; \theta) \\
\{\varepsilon \}_j &=& \varepsilon_j
\end{eqnarray*}
where $y_j(j=1, \cdots , m)$ are the measurement points at $\tau^*$, 
$U(\tau^*, y_j;\theta)$ solves the Cauchy problem (\ref{E-BS-consider}) for the unknown parameters $\theta$ and $\varepsilon_j$ is the uncertainty (noise) in the market, assumed as white Gaussian noise with 
a known standard deviation $\Sigma_{\varepsilon}$.
We then seek the parameters $\bar{\theta}$, 
which assumedly represent the true value of $\theta$, such that
\begin{equation}
Y_{\bar{\theta}} = F(\theta) + \varepsilon.
\label{Inverse main-model}
\end{equation}

\section{Bayesian inference approach to IOP}
The Bayesian  inference approach is now widely used with great successes for solving a variety of inverse problem (see for example \cite{mcmc-springer}). The solution of the Bayesian inference approach 
is estimated not as single-valued, 
but as the posterior conditional mean (CM)
\begin{equation}
\theta_{\bf CM}:=\int \theta \pi(\theta|Y) {\rm d} \theta,
\label{cm}
\end{equation}
of the unknown parameters $\theta$ given the measured data $Y$, where $\pi(\theta|Y)$ is the conditional density function, which is called the posterior probability density function (PPDF) in this study.
Now, according to the Bayes' theorem, PPDF is defined as follows:
\begin{equation}
\pi(\theta|Y) = \displaystyle \frac{\pi(Y|\theta)\pi(\theta)}{\pi(Y)}.
\label{Bayes'theorem}
\end{equation}
i.e. the posterior probability of a hypothesis is proportional to the product of its likelihood and its prior probability. The likelihood function $\pi(Y|\theta)$ is then given as
\begin{equation}
\pi(Y|\theta) = \exp\left\{ -\displaystyle \frac{(Y-F(\theta))^{T}(Y-F(\theta))}{2\Sigma^2_\varepsilon} \right\}.
\end{equation}
In some case, since we don't know much about a prior density function $(\theta) $, it is simply assumed as $\pi(\theta) = U_{[-\theta_0,\theta_0]}$, where $\theta_0$ is a sufficiently large positive constant. Thus, the PPDF of the parameters $\theta$ is the same as the likelihood function as follows (cf. \cite{mcmc-springer}):
\begin{equation}
\pi(\theta|Y) \propto  \exp\left\{ -\displaystyle \frac{(Y-F(\theta))^{T}(Y-F(\theta))}{2\Sigma^2_\varepsilon} \right\}.
\end{equation}

\subsection{MCMC methods}

It is hard to know the explicit form of $\pi(\theta|Y)$, Markov chain Monte Carlo (MCMC) algorithm given in Robert and Casella \cite{Robert-Casella} can be applied to obtain a set of samples $\theta_k$ ($k = 1,\cdots, K$) and these independent samples can approach the distribution $\pi(\theta|Y) $. Also the posterior conditional mean comes to
$$\theta_{\bf CM}\approx \frac{1}{K}\sum_{k=1}^K \theta_k.$$
This is the solution of our IOP under the meaning of statistics. 

In this paper, we employs a typical MCMC algorithm called the Metropolis--Hastings (M--H) algorithm (see Metropolis et al. \cite{Metropolis}; Hastings \cite{Hastings}). 
M--H Algorithm given below builds its Markov chain by accepting or rejecting samples extracted from a proposed distribution. M--H Algorithm is generally used in Bayesian inference approach (cf. \cite{mcmc-springer}).

\vspace{2ex}

\noindent\textbf{\underline{M--H Algorithm}}
\begin{itemize}
  \item{\textbf{Step1}: Generate $\theta'\sim q(\cdot|\theta_k)=N(\theta_k,\gamma^2)$ (the normal distribution) with a given stander derivation $\gamma>0$ for given $\theta_k$.}
  \item{\textbf{Step2}: Calculate the acceptance rate $\alpha(\theta',\theta_k)=\min\left\{1,f(\theta'|Y)/f(\theta_k|Y)\right\}$.}
  \item{\textbf{Step3}: Update $\theta_k$ as $\theta_{k+1}=\theta'$ with probability $\alpha(\theta',\theta_k)$ but otherwise set $\theta_{k+1}=\theta_k$ and re-sample from 1.}
\end{itemize}

While running this M--H algorithm, we can find, by given any initial guess $\theta_0$, the samples will come to a stable Markov chain after a burn-in time $k^*$. In other word, unlike common Newton--type iterative regularization methods 
(for example, the Levenberg--Marquardt (LM) algorithm), the MCMC algorithm does not highly depend on the initial guess and the mean value 
$$\theta_{\bf CM}\approx \frac{1}{K-k^*}\sum_{k=k^*+1}^K \theta_k,$$
always reaches the global minimum after a sufficiently long sampling time (cf. \cite{mcmc-springer}).

\section{Numerical examples}

In this section, 
we simultaneously estimate the unknown drift and volatility coefficients from the measured artificial data $Y_{\bar{\theta}}$ in (\ref{Inverse main-model}), by using a Bayesian inference approach. 
Here, in order to investigate a small perturbation of the drift around the interest $r$, 
we assume that the drift $\mu(y)$ has the following type:
\begin{equation}
\mu(y) = r + f(y),
\label{mu-form}
\end{equation}
where $f(y)$ is a small perturbation and $f(0)=0$.

\subsection{Direct problems}

In this section, we solve the direct problem for (\ref{E-BS-consider}) by the numerical Crank--Nicholson scheme:
\begin{eqnarray} 
\displaystyle -a_iU_{i+1,j+1}+\left(1+b\right)U_{i,j+1}-c_jU_{i-1,j+1}\hspace{15ex}\nonumber \\
\mbox{} \\
\hspace{10ex}=a_iU_{i+1,j}+\left(1-b-r\right)U_{i,j}+c_jU_{i-1,j},\nonumber 
\label{dis-eq}
\end{eqnarray}
where $U_{i,j}=U(y_i, \tau_j),$ and 
\begin{equation}
\displaystyle a_i=-\frac{\Delta\tau}{4\left(\Delta y\right)^2}
\left\{\sigma_0^2+\Delta y\left(\frac{1}{2}\sigma_0^2 - \mu(y_i) \right)\right\}, 
\end{equation}
\vspace{3mm}
\begin{equation}
\displaystyle b=\frac{\Delta\tau}{2\left(\Delta y\right)^2}\sigma_0^2, 
\end{equation}
\vspace{3mm}
\begin{equation}
\displaystyle c_i=-\frac{\Delta\tau}{4\left(\Delta y\right)^2}
\left\{\sigma_0^2 - \Delta y\left(\frac{1}{2}\sigma_0^2 - \mu(y_i)\right)\right\}. 
\end{equation}
Here, we took a uniform grid
\begin{eqnarray*}
\tilde\omega = \{(y_i, \tau_j): \ y_j \in {\boldmath I_{1.5}}=(-1.5,1.5), \ \tau_i \in (0, \tau^*), \hspace{12ex} \\
\hspace{30ex}
i=1,2,\cdots,100, j=1,2,\cdots,400 \} 
\end{eqnarray*}
with artificial Dirichlet boundary conditions at $y=-1.5$ and $1.5$, such as $U_{1,j}=1$ and $U_{100,j}=0$, and $\Delta y = y_{j+1}-y_j=\frac{1}{33}, \ \Delta \tau = \tau_{i+1}-\tau_i = 0.001$. 

Then (\ref{E-BS-consider}) can be given in the matrix form:
\begin{equation}
\displaystyle \mathbf{u}_{i+1}=
\mathbf{A}^{-1}\mathbf{B}\mathbf{u}_{i} -2 c_2 \mathbf{A}^{-1} \mathbf{e_{98}}, 
\vspace{3ex} 
\label{numeri-sol-U}
\end{equation}
where $\mathbf{u_i} = (U_{i,2}, U_{i,3}, \cdots, U_{i,99})^{T}$, 
$\mathbf{e_{98}} = (1, 0, \cdots, 0)^{T}$ and 
\begin{equation}
\mathbf{A}=\left(
  \begin{array}{cccccc}
    1+b   & -a_2   & 0   & 0   & \cdots    & 0  \\
    -c_3   & 1+b   & -a_3   & 0   & \cdots    & 0       \\
    0   & -c_4   & 1+b   & -a_4   & \cdots    & 0      \\
    \vdots    &    & \ddots    & \ddots    & \ddots    & \vdots        \\
    0   &    &    & -c_{98}   & 1+b   & -a_{98}       \\
    0   & \cdots    &    & 0   & -c_{99}   & 1+b       \\
  \end{array}
\right)_, 
\end{equation}
\vspace{3ex}

\begin{equation}
\mathbf{B}=\left(
  \begin{array}{cccccc}
    1-b-r   & a_2   & 0   & 0   & \cdots    & 0  \\
    c_3   &  1-b-r  & a_3   & 0   & \cdots    & 0       \\
    0   & c_4   & 1-b-r   & a_4   & \cdots    & 0      \\
    \vdots    &    & \ddots    & \ddots    & \ddots    & \vdots        \\
    0   &    &    & c_{98}   & 1-b-r   & a_{98}       \\
    0   & \cdots    &    & 0   & c_{99}   & 1-b-r       \\
  \end{array}
\right)_.
\end{equation}

\subsection{Inverse problem solution by MCMC}
\label{MCMC}

In the following examples, we set 
\begin{eqnarray}
\mu(y) = r + \theta_1 y + \theta_2 y^2 + \theta_3 y^3,
\end{eqnarray}
where $(\theta_1, \theta_2, \theta_3)$ are unknown constant parameters. 
Moreover the relative noise in all the measured artificial data $Y$ is assumed as $0\%$ and $5\%$, 
and the prior distribution $f(\theta)$ of unknowns is $$f_{\rm prior}(\theta)=U_{[\alpha^{\rm min},\alpha^{\rm max}]}(\alpha)\cdot U_{[\beta^{\rm min},\beta^{\rm max}]}(\beta)\cdot U_{[\gamma^{\rm min},\gamma^{\rm max}]}(\gamma)\cdot U_{[\sigma_0^{\rm min},\sigma_0^{\rm max}]}(\sigma_0)$$
and the intervals $[\alpha^{\rm min},\alpha^{\rm max}], [\beta^{\rm min},\beta^{\rm max}], [\gamma^{\rm min},\gamma^{\rm max}]$ and $[\sigma_0^{\rm min},\sigma_0^{\rm max}]$ are large enough. General uniform distributions can be used for $f(\theta)$ if we use the prior-reversible proposal that satisfies $f(\theta)q(\theta'|\theta)=f(\theta')q(\theta|\theta')$ (see for example \cite{Iglesias-Lin-Stuart14}). 
On the other hand, if we choose $f(\theta)$ as a Gaussian distribution, this will turn out to be the Tikhonov regularization term in the cost function (see also for example \cite{mcmc-springer}).

Then we simultaneously identify the parameters $(\theta_1, \theta_2, \theta_3)$ and $\sigma_0$ from the measured artificial data $Y_{\bar{\theta}}$.

For comparison, we particularly consider LM algorithm \cite{Levenberg44,Marquardt63}. That is, the  recovery of $\theta=(\theta_1, \theta_2, \theta_3, \sigma_0)^T$ is computed by the iteration given by
\begin{equation}
\theta_{k+1}=\theta_k+\left[F'(\theta_k)^TF'(\theta_k)+\lambda I\right]^{-1}F'(\theta_k)^T
\left(U-F(\theta_k)\right),
\label{LM:iteration}
\end{equation}
where $F'(a)$ is the Jacobian matrix and the parameter $\lambda$ is nonnegative. 
This algorithm can be implemented for example by an inner embedded program {\bf lsqcurvefit} in MATLAB $\circledR 2018 a$.

\vspace{2ex}

\noindent {\bf Example 1:}

In this example, we set 
\begin{eqnarray*}
f(y)=y,
\end{eqnarray*}
in the measured artificial data $Y_{\bar{\theta}}$. 
We seek parameters $\bar{\theta}=(\bar{\theta_1}, \bar{\theta_2}, \bar{\theta_3}, \bar{\sigma_0}) = (1,0,0,1)$ in this case.

\begin{figure}[h]
 \begin{center}
  \includegraphics[width=150mm]{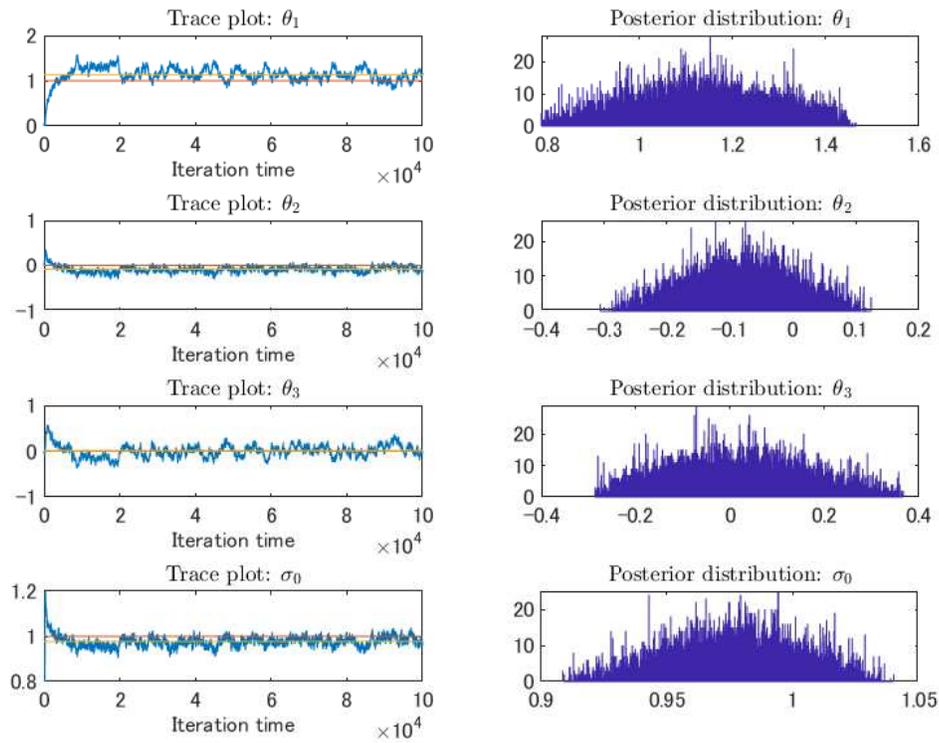}
  \caption{The trace plot and the posterior probability density function}
  \label{1para_ini000}
 \end{center}
\end{figure}

\begin{figure}[h]
 \begin{center}
  \includegraphics[width=100mm]{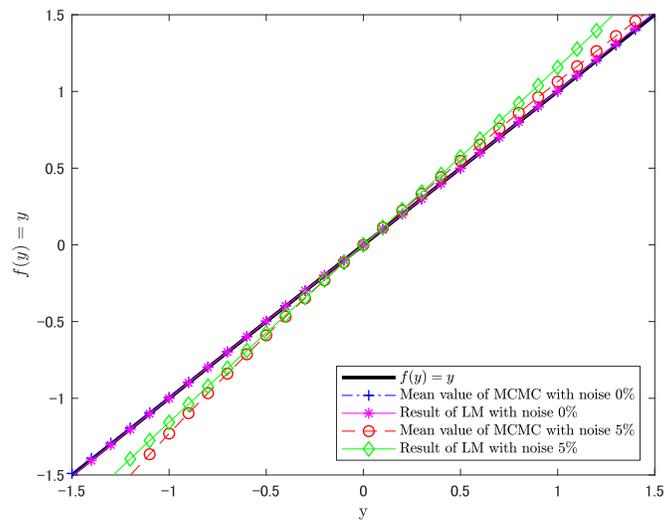}
  \caption{Recovered drift coefficient}
  \label{fitting_1para_ini000}
 \end{center}
\end{figure}

First, we set the initial guess of $(\theta_1, \theta_2, \theta_3, \sigma_0)$ as $(0,0,0,0)$ to the value far from the expected value $(1, 0, 0, 1)$. Figure \ref{1para_ini000} are the trace plots of the chain for $(\theta_1, \theta_2, \theta_3, \sigma_0)$ and PPDF. From results of trace plots, We can see that the chain mixes well.
Moreover, from these results and Table \ref{1para_ini000_table}, the recovery of $(\theta_1, \theta_2, \theta_3, \sigma_0)$ represents an excellent approximation of the expected value $(1, 0, 0, 1)$, even if the measured data $Y_{\bar{\theta}}$ is contaminated with noise.
Here, ``Mean value(with $0\%$ or $5\%$ noise)'' in Table \ref{1para_ini000_table} are the average of the value of the iteration time $100000$ after burn-in time $30000$.
For comparison, the converged recovery of $(\alpha, \beta, \gamma, \sigma_0)$ obtained by LM algorithm with $0\%$ and $5\%$ noise are also provided in Table \ref{1para_ini000_table}.
Moreover, from Figure \ref{fitting_1para_ini000}, 
we can see that the reconstruction of the drift is almost perfect near the center(i.e., at-the-money) in all cases.
In particular, the solution $f(y)$ obtained by MCMC is well reconstructed in the case of noise 0\%. 

\pagebreak

\begin{table}[h]
	\caption{Recovery results of $(\theta_1, \theta_2, \theta_3, \sigma_0)$.}
	\label{1para_ini000_table}
	\centering
	\tabcolsep=12pt
	\begin{tabular}{c||c|c|c|c}
		\hline
		\hline	
		Parameters & $\theta_1$ & $\theta_2$ & $\theta_3$ &	$\sigma_0$ \\
		\hline
		\hline
		Initial guess & 0 &0 & 0 & 0\\
		\hline
		Mean value(with $0\%$noise) &0.9966  & 0.0029 & 0.0005 & 1.0008\\
		\hline
		Result of LM with $0\%$noise & 0.9968 & 0.0020 & 0.0052 & 1.0003 \\
		\hline
		Mean value(with $5\%$ noise) &1.1346  & -0.0832 & 0.0121 & 0.9773\\
		\hline
		Result of LM $5\%$noise & 1.1434 & 0.0000 & 0.01357 &0.9836 \\
		\hline
		The expected value &1 & 0 & 0 & 1 \\
		\hline
		\hline
	\end{tabular}
\end{table}

\pagebreak

\begin{figure}[h]
 \begin{center}
  \includegraphics[width=150mm]{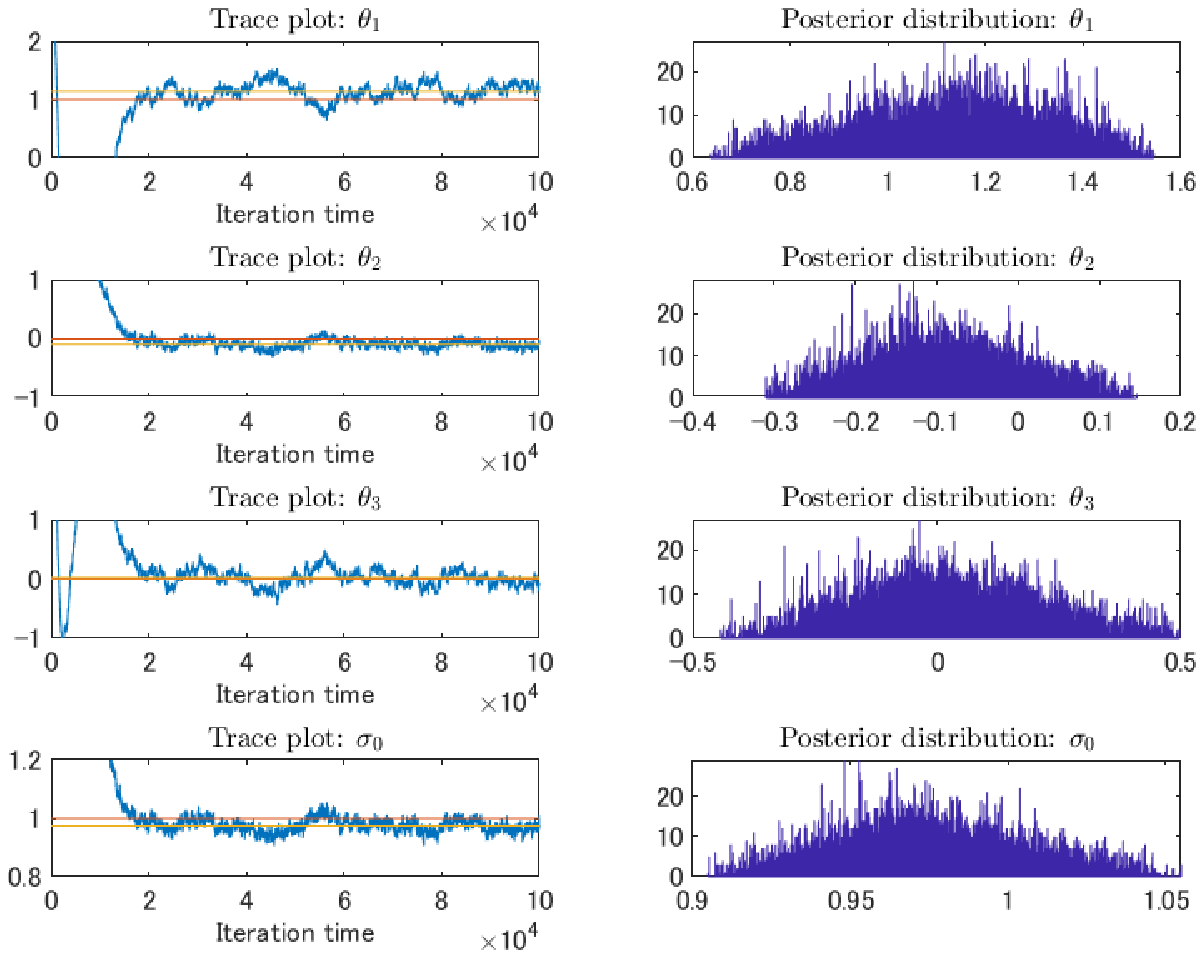}
  \caption{The trace plot and The posterior probability density function}
  \label{1para_ini035}
 \end{center}
\end{figure}

\begin{figure}[h]
 \begin{center}
  \includegraphics[width=100mm]{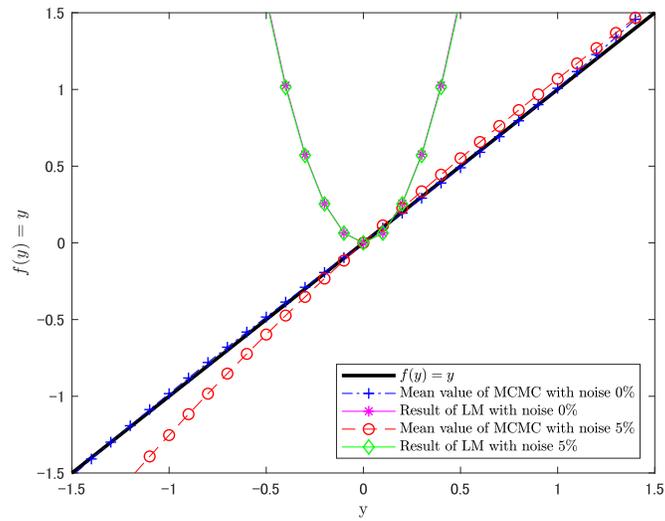}
  \caption{Recovered drift coefficient}
  \label{fitting_1para_ini035}
 \end{center}
\end{figure}

Next, the initial guess of $(\theta_1, \theta_2, \theta_3, \sigma_0)$ was set $(3.5,3.5,3.5,3.5)$ to the value far from the expected value $(1, 0, 0, 1)$. The evolutions of the MCMC sampled $\theta_1, \theta_2, \theta_3$ and $\sigma_0$ are shown in Figure \ref{1para_ini035}, and we can see that the chain mixes well.
Moreover recovered results for PPDF are presented in Figure \ref{1para_ini035}, and Table \ref{1para_ini035_table}.
From these results the recovery of $(\theta_1, \theta_2, \theta_3, \sigma_0)$ represents an excellent approximation of the expected value $(1, 0, 0, 1).$
The recovery of $(\theta_1, \theta_2, \theta_3, \sigma_0)$ obtained by the LM with $0\%$ and $5\%$ noise are also shown in Table \ref{1para_ini035_table}.
Moreover from Figure \ref{fitting_1para_ini035}, 
we can see that the reconstruction is almost perfect near the center except for the case of LM.

\pagebreak

\begin{table}[h]
	\caption{Recovery results of $(\theta_1, \theta_2, \theta_3, \sigma_0)$.}
	\label{1para_ini035_table}
	\centering
	\tabcolsep=12pt
	\begin{tabular}{c||c|c|c|c}
		\hline
		\hline	
		Parameters & $\theta_1$ & $\theta_2$ & $\theta_3$ &	$\sigma_0$ \\
		\hline
		\hline
		Initial guess & 3.5 &3.5 & 3.5 & 3.5 \\
		\hline
		Mean value(with $0\%$ noise) &0.9657  & 0.0124 & 0.0295 & 1.0045\\
		\hline
		Result of LM with $0\%$ noise & 0.0000 & 6.4108 & 0.0000 & 2.8517 \\
		\hline
		Mean value(with $5\%$ noise) &1.1451  & -0.0919 & 0.0159 & 0.9742\\
		\hline
		Result of LM with $5\%$ noise& 0.0000 & 6.3361 & 0.0000 & 2.8430 \\
		\hline
		The expected value &1 & 0 & 0 & 1 \\
		\hline
		\hline
	\end{tabular}
\end{table}

\pagebreak

In both the initial guess $(0, 0, 0, 0)$ and $(3.5, 3.5, 3.5, 3.5)$, from the results of the MCMC samples in Figure \ref{1para_ini000}, Figure \ref{1para_ini035} and the posterior condition mean values in Table \ref{1para_ini000_table}, \ref{1para_ini035_table} and Figure \ref{fitting_1para_ini000}, \ref{fitting_1para_ini035}, 
we can see that we succeeded in recovering parameters.

On the other hand, in the initial guess $(0, 0, 0, 0)$, the recoveries obtained by LM algorithm in Table \ref{1para_ini000_table} and Figure \ref{fitting_1para_ini000} succeeded as the case of MCMC algorithm.
However, 
in the initial guess $(3.5, 3.5, 3.5, 3.5)$, 
we could not obtain the results of the recovering parameters by LM algorithm in Table \ref{1para_ini035_table}, Figure \ref{fitting_1para_ini035}.
From these results, we observe that parameters are more sensitive to initial guess than MCMC algorithm and hence it is less easily recovered.

\pagebreak

\noindent {\bf Example 2:}

In this Example, we set 
\begin{eqnarray*}
f(y)=\sin y,
\end{eqnarray*}
in the measured artificial data $Y_{\bar{\theta}}$. 
In this case, 
we seek parameters $\bar{\theta}=(1,0,-1/6,1)$.
Here the expected values $(1, 0, -1/6)$ of $(\theta_1, \theta_2, \theta_3)$ are coefficients up to the third term of a Taylor series of the function $\sin y$ around $y = 0$.

\begin{figure}[h]
 \begin{center}
  \includegraphics[width=150mm]{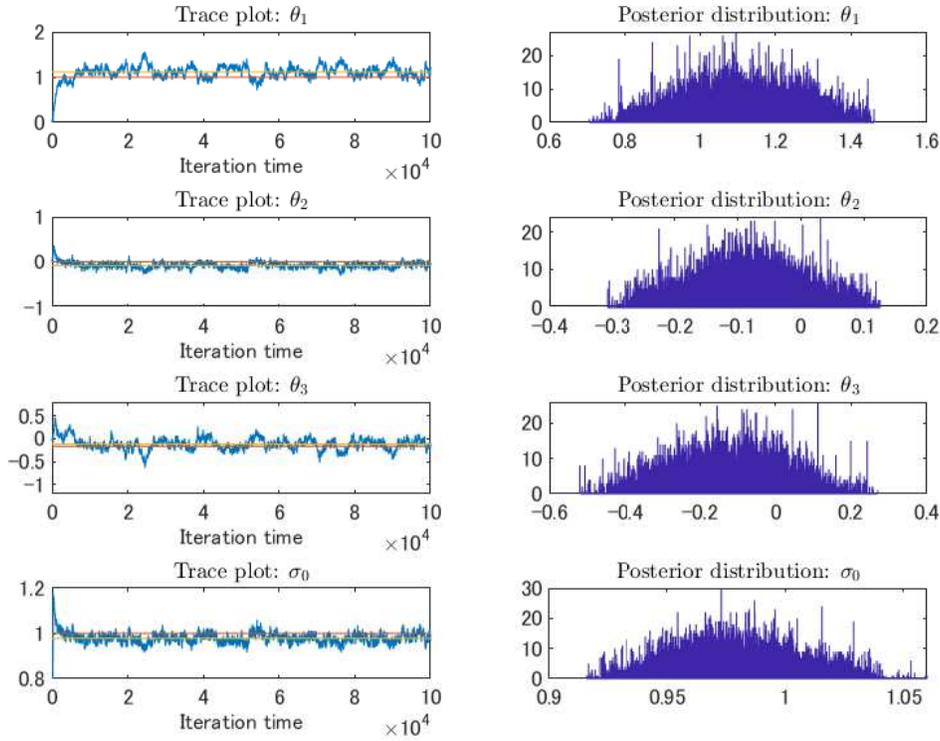}
  \caption{The trace plot and The posterior probability density function}
  \label{siny_ini000}
 \end{center}
\end{figure}

\begin{figure}[h]
 \begin{center}
  \includegraphics[width=100mm]{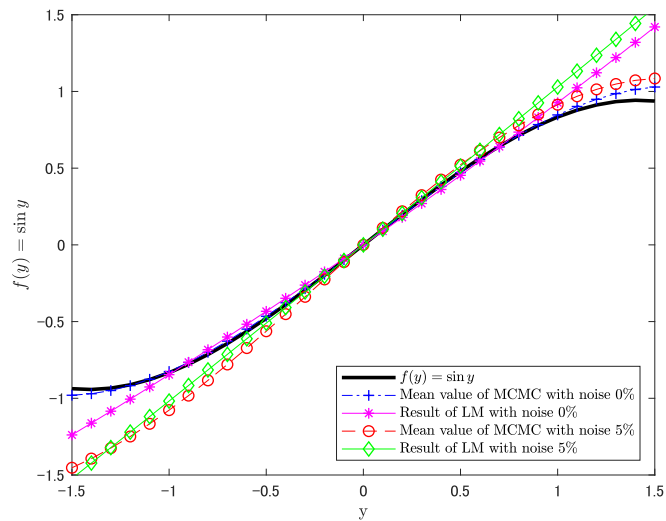}
  \caption{Recovered drift coefficient}
  \label{fitting_siny_ini000}
 \end{center}
\end{figure}

First, we set the initial guess of $(\theta_1, \theta_2, \theta_3, \sigma_0)$ as $(0,0,0,0)$ to the value far from the expected value $(1, 0, -1/6, 1)$. Figure \ref{siny_ini000} are the trace plots of the chain for $(\theta_1, \theta_2, \theta_3, \sigma_0)$ and the posterior probability density function.
From results of trace plots, we can see that the chain mixes well.
Moreover, from these results and Table \ref{siny_ini000_table}, the recovery of $(\theta_1, \theta_2, \theta_3, \sigma_0)$ represents an excellent approximation of the expected value $(1, 0, -1/6, 1)$, even if the measured artificial data $Y_{\bar{\theta}}$ is contaminated with noise.
For comparison, the converged recovery of $(\theta_1, \theta_2, \theta_3, \sigma_0)$ obtained by LM algorithm is also provided in Table \ref{siny_ini000_table}.
Further, from Figure \ref{fitting_siny_ini000}, we can see that results of all cases are almost perfect near the center. Especially, parameters obtained by MCMC with 0\% noise are recovered perfectly.

\begin{table}[h]
	\caption{Recovery results of $(\theta_1, \theta_2, \theta_3, \sigma_0)$.}
	\label{siny_ini000_table}
	\centering
	\tabcolsep=12pt
	\begin{tabular}{c||c|c|c|c}
		\hline
		\hline	
		Parameters & $\theta_1$ & $\theta_2$ & $\theta_3$ &	$\sigma_0$ \\
		\hline
		\hline
		Initial guess & 0 &0 & 0 & 0\\
		\hline
		Mean value(with $0\%$noise) &0.9694  & 0.0108 & -0.1331 & 1.0040\\
		\hline
		Result of LM with $0\%$noise & 0.8863 & 0.0404 & 0.0000 & 1.0148 \\
		\hline
		Mean value(with $5\%$ noise) &1.1151  & -0.0818 & -0.1197 & 0.9782\\
		\hline
		Result of LM with $5\%$ noise & 1.0234 & 0.0054 & 0.0000 &0.9946 \\
		\hline
		The expected value &1 & 0 & -1/6 & 1 \\
		\hline
		\hline
	\end{tabular}
\end{table}

\pagebreak

\begin{figure}[h]
 \begin{center}
  \includegraphics[width=150mm]{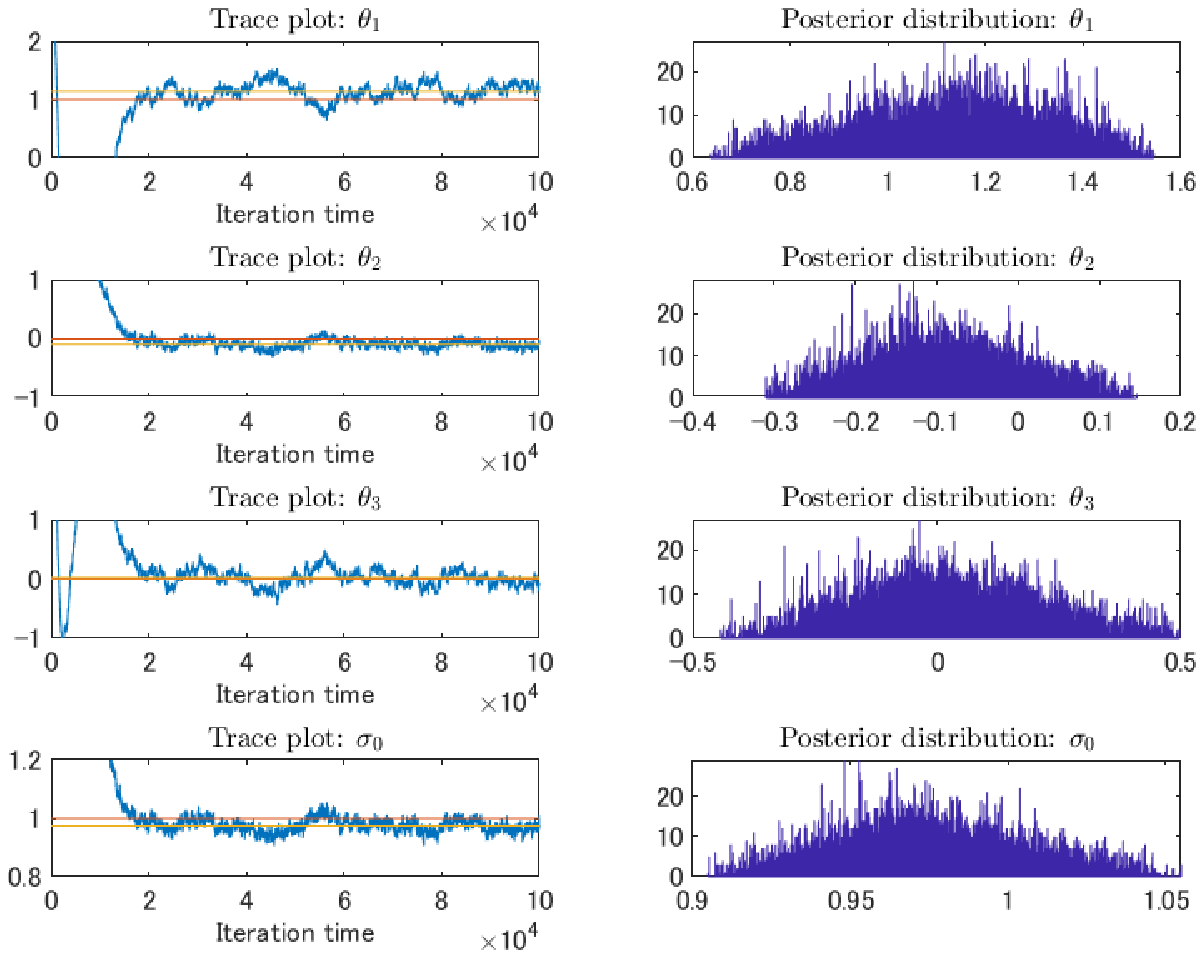}
  \caption{The trace plot and The posterior probability density function(PPDF)}
  \label{siny_ini035}
 \end{center}
\end{figure}

\begin{figure}[h]
 \begin{center}
  \includegraphics[width=100mm]{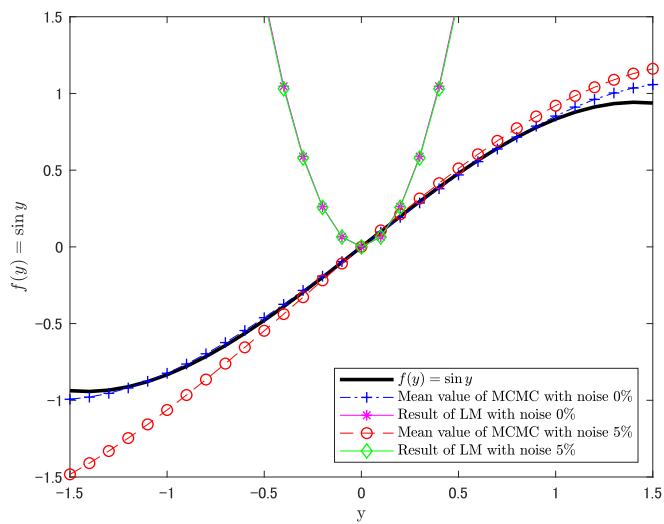}
  \caption{Recovered drift coefficient}
  \label{fitting_siny_ini035}
 \end{center}
\end{figure}

Next, the initial guess of $(\theta_1, \theta_2, \theta_3, \sigma_0)$ was set $(3.5,3.5,3.5,3.5)$ to the value far from the expected value $(1, 0, -1/6, 1)$. The evaluations of the MCMC sampled $\theta_1, \theta_2, \theta_3$ and $\sigma_0$ are shown in Figure \ref{siny_ini035}, and we can see that the chain mixes well.
Moreover recovered results for PPDF are presented in Figure \ref{siny_ini035}, and Table \ref{siny_ini035_table}.
From these results the recovery of $(\theta_1, \theta_2, \theta_3, \sigma_0)$ represents an excellent approximation of the expected value $(1, 0, -1/6, 1).$
The divergent recovery of $(\theta_1, \theta_2, \theta_3, \sigma_0)$ obtained by LM algorithm in Table \ref{siny_ini035_table}.
Moreover, from Figure \ref{fitting_siny_ini035}, 
we can see that the reconstruction is almost perfect near the center except for results of LM algorithm.
In particular, results of MCMC with 0\% noise are recovered perfectly.

\pagebreak

\begin{table}[h]
	\caption{Recovery results of $(\alpha, \beta, \gamma, \sigma_0)$.}
	\label{siny_ini035_table}
	\centering
	\tabcolsep=12pt
	\begin{tabular}{c||c|c|c|c}
		\hline
		\hline	
		Parameters & $\alpha$ & $\beta$ & $\gamma$ &	$\sigma_0$ \\
		\hline
		\hline
		Initial guess & 3.5 &3.5 & 3.5 & 3.5 \\
		\hline
		Mean value(with $0\%$ noise) &0.9617  & 0.0143 & -0.1235 & 1.0048\\
		\hline
		Result of LM with $0\%$ noise & 0.0000 & 6.5424 & 0.0000 & 2.8646 \\
		\hline
		Mean value(with $5\%$ noise) &1.0804  & -0.07152 & -0.0885 & 0.9824\\
		\hline
		Result of LM with $5\%$ noise & 0.0001 & 6.4377 & 0.0000 & 2.8553 \\
		\hline
		The expected value &1 & 0 & -1/6 & 1 \\
		\hline
		\hline
	\end{tabular}
\end{table}

\pagebreak

As in Example 1, in both the initial guess $(0, 0, 0, 0)$ and $(3.5, 3.5, 3.5, 3.5)$ from results of MCMC samples in Figure \ref{siny_ini000}, \ref{siny_ini035} and the posterior condition mean values in Table \ref{1para_ini000_table}, \ref{1para_ini035} and Figure \ref{fitting_1para_ini000}, \ref{fitting_1para_ini035}, 
we can see that we succeeded in recovering parameters. 

On the other hand, in the initial guess $(0, 0, 0, 0)$, 
the recoveries obtained by LM algorithm in Table \ref{siny_ini000_table} and \ref{fitting_siny_ini000} almost succeeded as the case of MCMC algorithm as in Example 1.
However, 
in the initial guess $(3.5, 3.5, 3.5, 3.5)$, we could not obtain the results of the recovering parameters by LM algorithm in Table \ref{siny_ini035_table} and \ref{fitting_siny_ini035}.
From this result, 
we observe that parameters are more sensitive 
to initial values than MCMC algorithm and hence it is less easily recovered as in Example 1.

\section{Conclusion}

In this study, we have established the method of simultaneous estimation of the unknown drift and volatility coefficients from the measured data, by using a Bayesian inference approach(MCMC-MH) based on a partial differential equation of parabolic type.
In particular, we took into account an application to real financial markets and dealt with the case with Heaviside function as the initial condition, so-called binary option.
In the instantaneous estimation of drift and volatility coefficients, 
we assumed that the volatility coefficient is a constant and the drift coefficient is a cubic function with three unknown parameters. 

The posterior distributions of the unknown drift and volatility coefficients were recovered from the measured data 
by modeling the measurement errors as Gaussian random variables.

The posterior state space was explored by the MCMC--M--H method.
As confirmed in the numerical results, 
the Bayesian inference approach(the MCMC algorithm) simultaneously estimated the unknown drift and volatility coefficients from the measured data than the Levenberg--Marquardt algorithm.

There are still several problems we have to settle.
First, from the form of our model it is expected that we will be able to apply the results of this study to problems of term structure models for an interest rate.
Moreover we will try to identify parameters of another financial model, for instance, such as the model including the dividend yield.
Next, we will develop mathematical results
(for instance, the uniqueness, stability, and existence) of IOP 
and extend our approach to two--dimensional Black--Scholes Equation such as "Spread Option".
Finally, we have to study how to apply our results to the real financial market, and repeat tests.

The first author would like to acknowledge the supports from JSPS Grant-in-Aid for Scientific Research (C) 18K03439.

\section*{Declarations}
\begin{itemize}
\item Funding

The first author would like to acknowledge the supports from JSPS Grant-in-Aid for Scientific Research (C) 18K03439.
\item Conflict of interest/Competing interests 

All authors declare that: (i) no support, financial or otherwise, has been received from any organization that may have an interest in the submitted work; and (ii) there are no other relationships or activities that could appear to have influenced the submitted work.

\item Availability of data and materials

The data that support the findings of this study are available from the corresponding author, [Yasushi Ota yasushio@andrew.ac.jp], upon reasonable request.

\end{itemize}

\bibliographystyle{unsrt}  
\bibliography{references}

\begin{thebibliography}{88}

\bibitem{Alif} Alifanov M. O. 1997 \textit{Inverse Heat Transfer Problems,} \textit{International Series in Heat and Mass Transfer,} Springer Verlag.

\bibitem{beck} Beck V. J., Blackwell B. and Clair Jr. C. R. 1985 \textit{Inverse Heat Conduction: Ill-Posed Problems,} Wiley-Interscience.

\bibitem{B-S} Black F. and Scholes M. 1973 \textit{The pricing of options and corporate liabilities,} \textit{Journal of Political Economy,} \textbf{81}, 637-659.

\bibitem{B-I} Bouchouev I. and Isakov V. 1999 \textit{Uniqueness, stability and numerical methods for the inverse problem that arises in financial markets,} \textit{Inverse Problems,} \textbf{15}, R95-R116.

\bibitem{B-I-V}
Bouchouev I., Isakov V. and Valdivia N. 2002 \textit{Recovery of volatility coefficient by linearization,} \textit{Quantitative Finance,} Vol2, 257-263.

\bibitem{B-G-R}
Bunnin F. O., Guo Y.  and Ren Y. 2002 \textit{Option pricing under model and parameter uncertainty using predictive densities} \textit{Statistics and Computing,} \textbf{12(1)}, 37--44.

\bibitem{Geothermal}
Cui T., Fox C., and O'fSullivan M.J. 2011 \textit{Bayesian calibration of a large-scale geothermal reservoir model by a new adaptive
delayed acceptance Metropolis Hastings algorithm,} \textit{Water Resource Research,} \textbf{47}, W10521.

\bibitem{dupire}
Dupire B. 1994 \textit{Pricing with a smil,} \textit{Risk,} 7 18-20.

\bibitem{Doi-yasu} Doi S. and Ota Y. 2018 \textit{Application of microlocal analysis to an inverse problem arising from financial markets} \textit{Inverse Problems.} \textbf{34}, N11.

\bibitem{Fri} Friedman A. 1983 \textit{Partial Differential Equations of Parabolic Type,} (Englewood Cliffs, N.J: Prentice-Hall).

\bibitem{Remote}
Haario H., Laine M., Lehtinen M., Saksman E. and Tamminen J. 2004
\textit{Markov chain Monte Carlo methods for high dimensional
inversion in remote sensing,} 
\textit{Journal of the Royal Statistical Society: Series B (Statistical Methodology),}
\textbf{66}, 591-608.

\bibitem{Hastings}
Hastings, W. 1970 \textit{Monte Carlo sampling methods using Markov chains and their application,} \textit{Biometrika,} \textbf{57}, 97-109.

\bibitem{Iglesias-Lin-Stuart14}
Iglesias M. A., Lin K and Stuart A. M. 2014 \textit{Well-posed Bayesian geometric inverse problems arising in subsurface flow,} \textit{Inverse Problems,} \textbf{30}, 114001.

\bibitem{J-J}
Jacquier E. and Jarrow R. 2000 \textit{Bayesian analysis of contingent claim model error,} \textit{Journal of Econometrics,} \textbf{94(1-2)}, 145--180.

\bibitem{J-P1}
Jacquier E. and Polson N. 2010 \textit{Bayesian methods in finance,} \textit{Oxford handbook of Bayesian econometrics,} 439--512 (Oxford University Press).

\bibitem{mcmc-springer} Kaipio J. and Somersalo E. 2005 \textit{Statistical and Computational Inverse Problems.} (New York: Springer)

\bibitem{kubo} Korolev M., Kubo H. and Yagola G. 2012 \textit{Parameter identification problem for a parabolic equation-application to the Black-Scholes option pricing model,} \textit{J. Inverse Ill-posed probl,} \textbf{20} No.3, 327-337.

\bibitem{Levenberg44}
Levenberg, K. 1944 \textit{A method for the solution of certain non-linear problems in least squares,} \textit{Quarterly Appl. Math.,} {\bf 2}, 164--168.

\bibitem{L-Y} Lishang J. and Youshan T. 2001 \textit{Identifying the volatility of underlying assets from option prices,} \textit{Inverse Problems,} \textbf{17}, 137-155. 

\bibitem{Marquardt63}
Marquardt D. 1963 \textit{An algorithm for least-squares estimation of nonlinear parameters,} \textit {SIAM J. Appl. Math.,} \textbf{11}, 431--441.

\bibitem{Seismic}
Martin J., Wilcox L.C., Burstedde C. and Ghattas O. 2012
\textit{A stochastic Newton MCMC method for large-scale statistical inverse
problems with application to seismic inversion,} 
\textit{SIAM Journal on Scientific Computing,}
\textbf{34(3)}, A1460-A1487.

\bibitem{Metropolis}
Metropolis N., Rosenbluth A., Rosenbluth M., Teller A. and Teller E. 1953 
\textit{Equations of state calculations by fast computing machines,} 
\textit{J. Chem. Phys.,} \textbf{21(6)}, 1087-1092.

\bibitem{miyasu} Mitsuhiro M. and Ota Y. 2015 \textit{Recovery of Foreign Interest Rates from Exchange Binary Options,} \textit{Computer Technology and Application,} \textbf{6}, 76-88.

\bibitem{yasukaji} Ota Y. and Kaji S. 2016 \textit{Reconstruction of local volatility for the binary option model,} \textit{J. Inverse Ill-posed probl.,} \textbf{24}, No.6 727-742. 

\bibitem{yasu_jiang_gen_ue} Ota Y., Jiang Y., Nakamura G. and 
Uesaka M. 2019 \textit{Bayesian inference approach to inverse problems in a financial mathematical model,} \textit{International Journal of Computer Mathematics,} submitted. 

\bibitem{Robert-Casella}
Robert, C. and Casella, G. 2004
\textit{Monte Carlo Statistical Methods.} (Springer Texts in Statistics)

\bibitem{Tunaru}
Tunaru R. 2015 \textit{Model risk in financial markets: From financial engineering to risk management,} World Scientific Pub Co Inc.

\bibitem{T-Z}
Tunaru R. and Zheng T. 2017 \textit{Parameter estimation risk in asset pricing and risk management: A Bayesian approach,} \textit{International Review of Financial Analysis,} \textbf{53}, 80--93.

\bibitem{Heat1}
Wang J. and Zabaras N. 2004 \textit{A Bayesian inference approach to the inverse heat conduction problem,} \textit{International Journal of Heat and Mass Transfer,} \textbf{47}, Issues 17-18, 3927-3941.

\bibitem{Heat2}
Wang J and Zabaras N. 2005 \textit{Hierarchical Bayesian models for inverse problems in heat conduction,} \textit{Inverse Problems,}
\textbf{21} 183-206.

\end{thebibliography}

\end{document}